\documentclass[aps,prb,twocolumn,groupedaddress,showpacs]{revtex4}
\usepackage{graphicx}
\usepackage{dcolumn}
\usepackage{bm}

\begin{document}


\title{Evidence of local effects in anomalous refraction and focusing
properties of dodecagonal photonic quasicrystals}



\author{Emiliano Di Gennaro}
\email[]{emiliano@na.infn.it}
\affiliation{CNISM and Department of Physics, University of Naples ``Federico II'', Piazzale Tecchio 80, I-80125 Naples, ITALY}

\author{Carlo Miletto}
\affiliation{CNISM and Department of Physics, University of Naples ``Federico II'', Piazzale Tecchio 80, I-80125 Naples, ITALY}

\author{Salvatore Savo}
\affiliation{CNISM and Department of Physics, University of Naples ``Federico II'', Piazzale Tecchio 80, I-80125 Naples, ITALY}

\author{Antonello Andreone}
\affiliation{CNISM and Department of Physics, University of Naples ``Federico II'', Piazzale Tecchio 80, I-80125 Naples, ITALY}

\author{Davide Morello}
\affiliation{Waves Group, Department of Engineering, University of Sannio, Corso
Garibaldi 107, I-82100 Benevento, ITALY}

\author{Vincenzo Galdi}
\affiliation{Waves Group, Department of Engineering, University of Sannio, Corso
Garibaldi 107, I-82100 Benevento, ITALY}

\author{Giuseppe Castaldi}
\affiliation{Waves Group, Department of Engineering, University of Sannio, Corso
Garibaldi 107, I-82100 Benevento, ITALY}

\author{Vincenzo Pierro}
\affiliation{Waves Group, Department of Engineering, University of Sannio, Corso
Garibaldi 107, I-82100 Benevento, ITALY}

\date{\today}

\begin{abstract}
We present the key results from a comprehensive study of the
refraction and focusing properties of a two-dimensional dodecagonal
photonic ``quasicrystal'' (PQC), carried out via both full-wave
numerical simulations and microwave measurements on a slab made of
alumina rods inserted in a parallel-plate waveguide. We observe
anomalous refraction and focusing in several frequency regions,
confirming some recently published results. However, our
interpretation, based on numerical and experimental evidence,
differs substantially from the one in terms of ``effective negative
refractive-index'' that was originally proposed. Instead, our study
highlights the critical role played by short-range interactions
associated with local order and symmetry.
\end{abstract}


\maketitle Since the pioneering work by Yablonovitch
\cite{Yablonovitch} and John \cite{John}, ``photonic crystals''
(PCs) have elicited great attention from the scientific community,
in view of the variety of peculiar electromagnetic (EM)
 bandgap, waveguiding/confinement, refraction, and emission effects
attainable through their use. Among the most intriguing
applications, it is worth mentioning those to negative refraction
and subwavelength imaging (``superlensing'')
\cite{Notomi,Soukoulis,Parimi,Cubukcu}. The most typical PC
configurations are based on dielectric inclusions (or voids) arranged
according to {\em periodic} lattices in a host medium, and can
thus be studied using well-established tools and concepts such as
Bloch theorem, unit cell, Brillouin zone, equifrequency surfaces,
etc.

With specific reference to lensing applications, two different
approaches have been presented to obtain subwavelength resolution
using a dielectric PC slab. In the first one, a PC with high
dielectric contrast is tuned so as to behave (usually near a frequency band edge)
like a homogeneous material with a negative refractive index $n =
-1$ \cite{Notomi}, and the focus position of the flat lens follows
a simple ray-optical construction \cite{Lu}. In the second
approach, ``all angle negative refraction'' (AANR) is achieved
without an effective negative index, provided that the equifrequency
surfaces (EFSs) of the PC are all convex and larger than the
one pertaining to the host medium \cite{Luo}. In this
case, the focus position does not follow the ray-optical
construction and is {\em restricted} \cite{Wang}.

During the last decade, the discovery in solid-state physics of
certain metallic alloys (the so-called ``quasicrystals''
\cite{Shechtman,Levine}) whose X-ray diffraction spectra exhibit
``noncrystallographic'' rotational symmetries (e.g., 5-fold or
($K>6$)-fold, known to be incompatible with spatial periodicity) has
generated a growing interest toward {\em aperiodically-ordered}
geometries, leading to the study of the so-called ``photonic
quasicrystals'' (PQCs). In this framework, useful tools for
geometrical parameterization can be borrowed from the theory of
``aperiodic tilings'' \cite{Senechal}. Several recent numerical and
experimental studies have explored the EM properties of PQCs, in the
form of two-dimensional (2-D) aperiodic arrays of cylindrical rods
or holes, as well as 3-D structures fabricated via stereolithography
(see \onlinecite{PQC_review} and the references therein for a recent
review of the subject).

The study of PQCs entails significant complications, from both
theoretical and computational viewpoints, as compared to standard
(periodic) PCs. In spite of the lack of the aforementioned
Bloch-type concepts and tools, approaches to the calculation of the
density of states in PCQs have been proposed, relying, e.g., on
rational approximants \cite{Carlsson, David} or on extended zone
schemes in the reciprocal space \cite{Kalit2}. However, many PQC
properties and underlying mechanisms, which generally involve short-
and long-range interactions as well as complex multiple scattering
phenomena, are not yet fully understood. Nevertheless, results have
revealed the possibility of obtaining similar properties as those
exhibited by periodic PCs, with interesting potentials (e.g., richer
bandgap structure with lower and/or multiple frequencies of
operation, higher isotropy, easier achievement of phase-matching
conditions, etc.) in view of the additional degrees of freedom
typically available in aperiodically-ordered structures.

Recently, some examples of 2-D PQCs characterized by high-order
(8-fold, 10-fold, and 12-fold) rotational symmetries have been
proposed as good candidates to exhibit negative refraction and
subwavelength focusing effects with polarization-insensitive and
non-near-field imaging capabilities \cite{Feng, Zhang2}. Similar
results were also obtained in case of acoustic waves \cite{Zhang3}.
Such effects and properties were interpreted in \onlinecite{Feng}
within the framework of ``effective negative refractive-index and
evanescent wave amplification.''

In this letter, we present the key results from a comprehensive
study, based on full-wave numerical simulations backed by
experimental verifications, which show that the ``effective negative
refractive-index'' interpretation is questionable and unable to
fully explain and predict the above effects, which instead arise
from complex near-field scattering effects and short-range
interactions critically associated to local symmetry points in the
PQC.
%
\begin{figure}
\includegraphics[width=8cm]{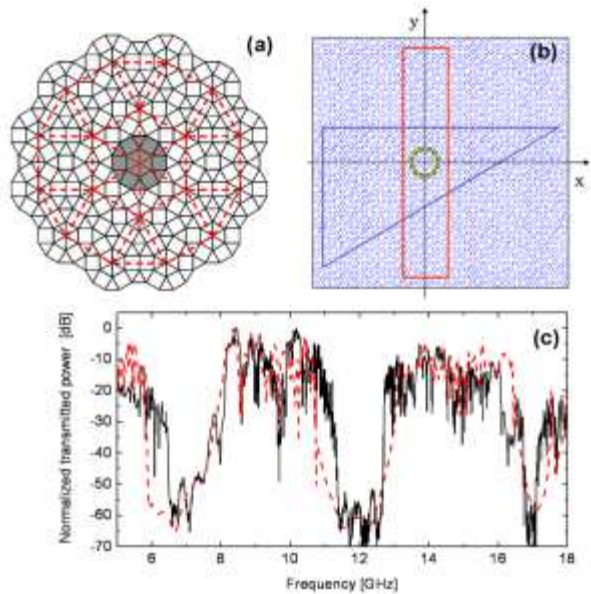}
\caption{\label{Figure1}(color online) (a): Illustration of the
Stampfli inflation rule. Starting from the parent tiling represented
by  the gray-shaded central dodecagon, a big parent (red dashed
lines) is generated by inflation, and filled up with copies of the
original dodecagon placed at its vertices. (b): A portion of the
tiling with the slab ($\sim7a$ thickness) and wedge realizations
considered in \onlinecite{Feng} (as well as in our study) marked
 by red and blue solid contours, respectively. The green-full-dot
dodecagon corresponds to the parent tiling shown in (a). (c):
Comparison (in a normalized scale) between the measured (black
continuous curve) and simulated (red dashed curve) transmitted power
for the PQC slab in (b) with $a=1.33$cm.}
 \end{figure}

The PQC of interest is a conformally-scaled version of that in
\onlinecite{Feng}, made of dielectric alumina rods (relative
permittivity $\epsilon_r$ = 8.6) of radius $r=0.4$ cm placed (in
air) at the vertices of a square-triangle 12-fold-symmetric tiling
with lattice constant (tile sidelength) $a = 1.33$ cm, generated
according to the Stampfli recursive construction \cite{Oxborrow}
(see also Fig. \ref{Figure1}a). Figure \ref{Figure1}b illustrates
two particular slab- and wedge-shaped tiling cuts (as in
\onlinecite{Feng}) considered in our study.

In our numerical simulations, assuming the rods infinitely long and
parallel to the electric field, we employ a well-established
full-wave technique (based on Bessel-Fourier multipolar expansion
\cite{Felbacq}), which has been extensively applied to the study of
2-D finite-size PCs \cite {Asatryan} and PQCs \cite {DellaVilla}.
The experimental verification relies on microwave (X-band)
measurements on PCQ slabs made of alumina rods of height $h=1$cm
inserted in an aluminum parallel-plate waveguide terminated with
microwave absorbers, with a monopole antenna used to generate an
electric field parallel to the rods. The intensity/phase maps are
collected using a HP8720C Vector Network Analyzer and a
computer-controlled $x-y$ movable monopole antenna probe, in a setup
similar to that described in \onlinecite{Parimi,Parimi2}.

As a preliminary check, we studied the transmission properties of
the PQC slab in Fig. \ref{Figure1}b, for a fixed source position,
within the $5-18$GHz frequency range. Figure \ref{Figure1}c compares
the experimental and simulated (normalized) transmitted power
through the slab. Three main bandgap regions are clearly observed,
with fairly good agreement between simulations and measurements. We
then studied the focusing properties within the frequency region
$8-10$ GHz (between the first and second bandgaps), by observing the
field intensity maps at the image side. We found several frequency
regions (including those reported in \onlinecite{Feng}) where a
clear focus was visible, with bandwidth ranging from few to
thousands MHz. We note that this feature represents a first
remarkable difference with the (periodic) PC case, where the
focusing regions tend to be more rare and well separated. The
seemingly {\em denser} occurrence of focusing effects in PQCs could
be attributed to their inherent self-similar nature \cite{Kalit}. We
observed the most stable focusing effects around $8.350$ GHz.
However, in what follows we concentrate on the focusing effects at
$8.836$ GHz, corresponding to the configuration considered in
\onlinecite{Feng}. In our study emphasis is not placed on {\em
quantitative} assessments of the subwavelength focusing
capabilities, but rather on the phenomenology interpretation, and
its similarities and differences with respect to the periodic PC
case.
%
\begin{figure}
\includegraphics [width=8cm]{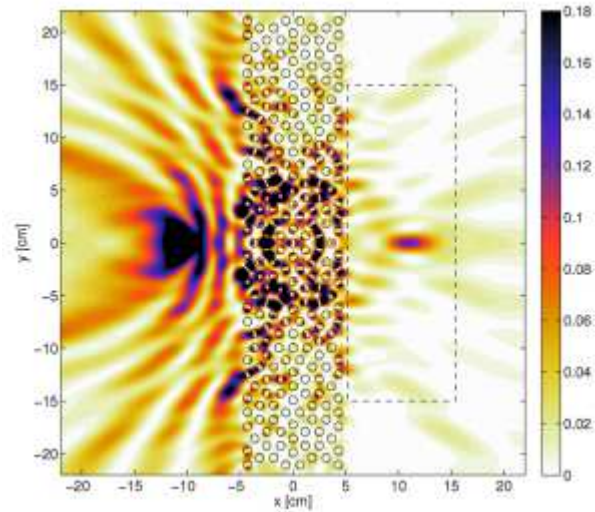}
\caption {\label{Figure2}(color online) Simulated field intensity
map at 8.836 GHz for a PQC slab as in Fig. \ref{Figure1}c (with
lateral width of $42.9$ cm and thickness of $9.4$ cm), illustrating
the focusing of a source placed at $y_s=0$ and at a distance
$d_x=6$cm from the slab surface. The dashed rectangle
 delimits the 10cm$\times$30cm area scanned in the image-side measurements.}
\end{figure}

Figure \ref{Figure2} shows an example of a field intensity map
inside and outside the PQC slab in Fig. \ref{Figure1}b. It is worth
noticing that such slab is centered at the center of the tiling
($x=y=0$, see Fig. \ref{Figure1}b), which is not only a center of
local (12-fold) rotational symmetry, but also possesses reflection
symmetries with respect to both $x$ and $y$ axes. The presence of a
focus at the image side is clearly observed. However, from a
comprehensive numerical parametric study (see also \onlinecite{pnfa}
for more details), we found that neither a ray diagram approach can
be used nor a preferential propagation direction can be established
to justify and predict the image formation. Conversely, we found
that a local structure (in particular, the green-full-dot dodecagon
parent tiling highlighted in Fig. \ref{Figure1}b) around the local
symmetry center plays a key role.

If the PQC slab behaved like a homogeneous material with a negative
refractive index $n=-1$, the source-image distance would remain
constant and equal to twice the slab thickness \cite{Lu}. Results
for periodic PCs showing an almost isotropic refractive response
(see, e.g., \onlinecite{Wang}) agree fairly well with this
prediction. Moreover, in PC lenses, for lateral displacements of the
source that preserve the distance from the lens interface as well as
the structure periodicity, the focus remains unaffected, and its
position follows the source location, in view of the absence of an
optical axis.
%
\begin{figure}
\includegraphics [width=8cm]{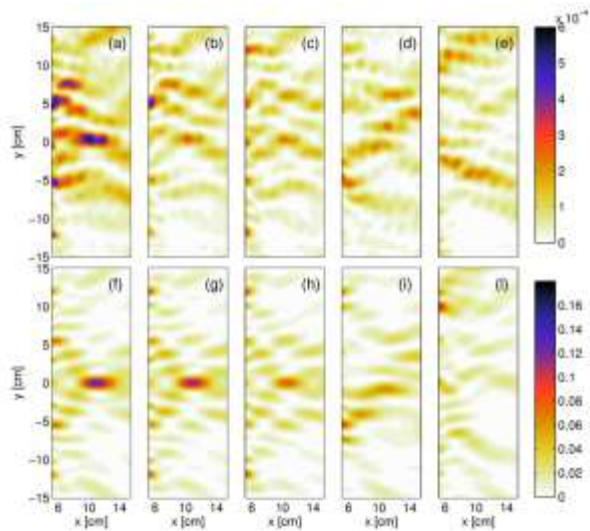}
\caption {\label{Figure3}(color online) As in Fig. \ref{Figure2},
but details of the measured (a--e) and simulated (f--l) field
intensity maps at the image side, for various source positions. (a),
(f): Source at $y_s=0$, and at a distance $d_x=6$cm from the slab
surface (cf. Fig. \ref{Figure2}); (b), (g): $y_s=0$ and $d_x=7$cm;
(c), (h): $y_s=0$ and $d_x=8$cm; (d), (i): $y_s=1$cm and $d_x=6$cm;
(e), (l): $y_s=5$cm and $d_x=6$cm.}
 \end{figure}

Figure \ref{Figure3} shows some representative measured and
simulated field intensity maps at the image side of the PQC slab,
for orthogonal and parallel (to the slab interface) displacements of
the source. Again, a general good agreement is observed between
simulations and measurements. Specifically, Figs. \ref{Figure3}a and
\ref{Figure3}f pertain to the configuration in Fig. \ref{Figure2},
whereas Figs. \ref{Figure3}b, \ref{Figure3}g and Figs.
\ref{Figure3}c, \ref{Figure3}h pertain to source displacements along
the $x-$axis (i.e., orthogonal to the slab interface), which
preserve the $y_s=0$ position (i.e., keeping the source facing the
local symmetry center of the tiling). The focus position does not
change substantially, raising further concerns about the
interpretation in \onlinecite{Feng}. Moving the source parallel to
the slab surface, and therefore breaking the symmetry around the
$x-$axis, one observes from Figs. \ref{Figure3}d, \ref{Figure3}i and
Figs. \ref{Figure3}e, \ref{Figure3}l that even small displacements
significantly affect the focus image, which undergoes a rapid
deterioration until it completely disappears for the source placed
at $y_s=5$cm. Interestingly, a focus can be still observed as long
as one extends the slab size along the $y-$axis and places the
source at $y_s=13.67$cm, i.e., directly facing the symmetry center
of another big parent tiling (see Fig.\ref{Figure1}b). In this case,
however, the image (not shown for brevity) exhibits a worse quality.
The complex interplay between local and global order and symmetry in
the focusing properties, which can be glanced from the above
results, is confirmed by parametric studies of PQC slabs of
different thicknesses, where a variety of effects can be observed,
ranging from localized (single and multiple) spots to beaming
phenomena (see \onlinecite{pnfa} for details).
%
\begin{figure}
\includegraphics [width=8cm]{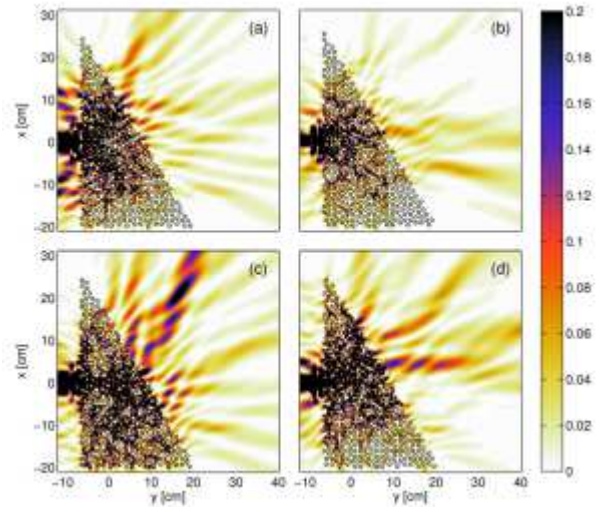}
\caption {\label{Figure4}(color online) Simulated intensity field
maps at $8.836$ GHz for a collimated Gaussian beam (with minimum
spot-size $\sim 2.8$cm) normally impinging on several realizations
of a PQC $60^o$-wedge. (a): Incident beam axis intersecting the
local symmetry center, as in \onlinecite{Feng}[Fig. 2]. (b)--(d):
Different wedge realizations obtained by random rigid translations
of the tiling (so as to displace the local symmetry center from the
incident beam axis).}
\end{figure}

As a further check, we also carried out a numerical study involving
a collimated Gaussian beam impinging with normal incidence on the
surface of PQC $60^o$-wedges extracted from the dodecagonal tiling.
Results are shown in Fig. \ref{Figure4}. Specifically, Fig.
\ref{Figure4}a pertains to the wedge realization shown in Fig.
\ref{Figure1}b, with the incident beam axis intersecting the local
symmetry center. Again, our results are similar to those in
\onlinecite{Feng}, with the transmitted beam  propagating mainly in
the ``negative'' direction -- a phenomenon previously interpreted
within the framework of ``effective negative refractive-index.''
However, a deeper study reveals that this effect too is critically
related to the mutual position of the incident beam and the local
symmetry center. This is clearly visible in Figs.
\ref{Figure4}b--\ref{Figure4}d, pertaining to different PQC wedge
samples obtained by random rigid translations of the tiling (so as
to displace the local symmetry center from the incident beam axis),
which display complex multi-beam features in the transmitted field,
thereby highlighting the absence of a clear-cut refractive behavior.
%
\begin{figure}
 \includegraphics[width=8cm]{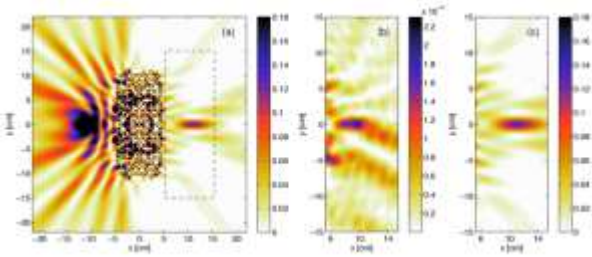}
\caption {\label{Figure5}(color online) (a): As in Fig.
\ref{Figure2}, but for a PQC slab with lateral width reduced to
$22$cm. (b), (c): Details of the measured and simulated field
intensity maps, respectively, at the image side.}
\end{figure}

From the above results, which confirm the key role played by {\em
short range} interactions involving a neighborhood of the parent
tiling, one could speculate that the focusing properties of a PQC
slab would be restricted to a limited range of incidence angles, and
should therefore occur also for significantly reduced lateral
widths. Indeed, the focusing effects turn out to be rather {\em
robust} with respect to lateral width reductions that do not affect
significantly the modal field distribution in a neighboring region
of the local symmetry center. Figure \ref{Figure5} shows the
simulated and experimental results pertaining to a PQC slab with
lateral width of only $22$cm (i.e., $\sim$ six wavelengths, nearly
half of that in Fig. \ref{Figure2}), and yet still exhibiting a
clear focus. Similar results were also obtained for PQC slabs with
even smaller (only three wavelengths) lateral width, but with larger
thickness \cite{pnfa}.

In conclusion, our numerical and experimental study of the
refraction and focusing properties of dodecagonal PQCs confirms some
of the results reported in the recent literature \cite{Feng}, but
shows that, contrary to the original interpretation, such results
are not attributable to an ``effective negative refractive-index.''
Instead, they arise from complex near-field scattering effects and
short-range interactions critically associated to local symmetry
points in the PQC, which were only glossed over in previous studies.
In this connection, it is worth recalling that local order and
symmetry have already been observed to play a key role in a variety
of PQC-related effects, including bandgap formation
\cite{DellaVilla}, field localization \cite{DellaVilla2}, and
directive emission \cite{Micco}. The new evidence and insights
presented in this letter constitute a further step toward a full
understanding of the underlying mechanisms, which remains crucial
for their judicious exploitation in the design of novel {\em
compact} optical systems. Within this framework, further numerical
and experimental studies of various PQC geometries are worth
pursuing.

\begin{acknowledgments}
This work has been funded by the Italian Ministry of Education and
Scientific Research (MIUR) under the PRIN-2006 grant on ``Study
and realization of metamaterials for electronics and TLC
applications.''
\end{acknowledgments}

\end{document}